# On the "mystery" of differential negative resistance

**Sebastian Popescu, Erzilia Lozneanu** and **Mircea Sanduloviciu**
Department of Plasma Physics
Complexity Science Group
Al. I. Cuza University
6600 Iasi, Romania
seba@uaic.ro

Investigating the causes of the nonlinear behavior of a gaseous conductor we identified the presence of two states of the complex space charge configuration self-assembled in front of the anode. These states correspond to two levels of self-organization from which the first one is related to spatial pattern whereas the second one to spatiotemporal pattern. Their emergence through two kinds of instabilities produced for two critical distances from thermodynamic equilibrium is emphasized in the current voltage characteristic as an S-shaped, respectively Z-shaped bistability. Their presence attributes to the gaseous conductor the ability to work as an S-shaped, respectively an N-shaped negative differential resistance.

## 1. Introduction

Many physical, chemical or biological systems driven away from thermodynamical equilibrium may give birth to a rich variety of patterns displaying strong structural and dynamical resemblances. The study of these nonlinear structures reveals many interesting features as multiplicities of states, hysteresis, oscillations and eventually chaos. In a characteristic in which a response parameter of the analyzed system is represented as a function of a control parameter, the multiplicity of states (*e.g.* bistabilities) is intimately related to the hysteresis phenomenon proving that the system has memory [1].

Informative for the genuine origin of the memory effect are experimental investigations recently performed on gaseous conductors [2,3]. Thus, when the gaseous conductor (plasma) is gradually driven away from the thermodynamic equilibrium the system reveals two different bistable regions (bifurcations) that successively appear in the form of an S-shaped, respectively an N-shaped **n**egative **d**ifferential **r**esistance (NDR). Their appearance emphasizes the development of two successive levels of self-organization [2,3].

The aim of the present paper is to show that the S-shaped DNR is related to a first level of self-organization, the final product of which is a stable (static) **c**omplex **s**pace

**c**harge **c**onfiguration (CSCC). The stability of CSCC is ensured by an electrical **d**ouble **l**ayer (DL) at its boundary. The sensitive dependence on the current of the emergence and deaggregation of the stable CSCC attributes to the gaseous conductor, in which such a self-organization structure appears, the quality of an S-shaped NDR. When the distance from the equilibrium is further increased by increasing the inflow of matter and energy, the CSCC transits into a new, higher level of self-organization. This is a steady state of the CSCC during which the DL from its border sustains and controls a rhythmic exchange of matter and energy between the complexity and the surrounding environment.

The presented experimental results offer, in our opinion, a new insight into a phenomenology essentially based on a self-organization scenario, the consideration of which could substantially advance the theoretical basis of a class of phenomena frequently classified as mysteries of Physics.

## 2. Levels of self-organization at the origin of S-, respectively Z-shaped bistabilities

The bistability type of a nonlinear system is usually related to different electrical transport mechanisms illustrated in the shape of the current-voltage characteristics. Thus the S-type characteristic specific to pnpn-diodes, quantum-dot structures etc. are related to a transport mechanism that involves impact-ionization breakdown, whereas the Z-shape characteristic is currently associated to moving domains, a mechanism observed in Gunn diodes, p-germanium etc. The fact that distinct shapes of the current-voltage characteristics were observed when the nonlinearity of different systems was investigated has led to the opinion that the physical processes at their origin are related to different basic phenomena, two of them mentioned above. That this is not so will be illustrated on the example of a gaseous conductor, the current-voltage characteristic of which reveals, when the anode voltage is gradually increased, the successive appearance of the S-shaped, respectively of the Z-shaped bistabilities. The cause of their appearance is an evolution mechanism during which a spatial pattern (stable state of the CSCC) emerges first in the gaseous conductor and, when the voltage of the dc power supply is gradually increased, spatiotemporal patterns appears (steady state of the CSCC). These levels of self-organization evolve through two instabilities by a mechanism whereby the first level favors further evolution by increasing the nonlinearity and the distance from equilibrium.

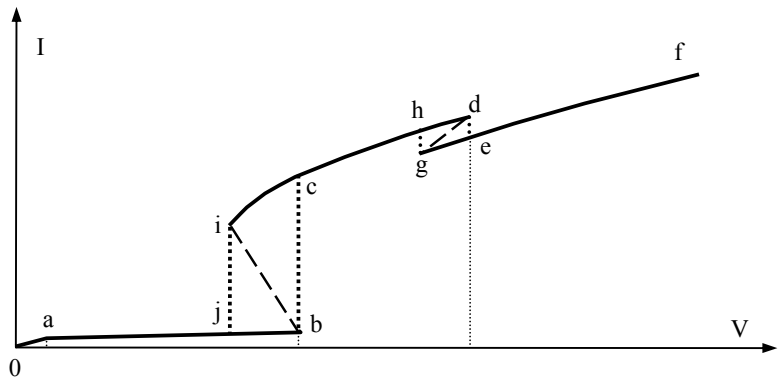

**Figure 1.** Typical static current *versus* voltage characteristic for a diode-like plasma system. The solid and dashed lines denote the stable, respectively unstable branches, and the dotted ones mark jumps.

The aim of our paper is to demonstrate that these bistabilities are related to a hierarchy of self-organization phenomena, the genesis of which involve key processes as symmetry breaking, bifurcations and long-range order *i.e.,* concepts at present considered as basic for the science of complexities.

The static current-voltage characteristic for a plasma diode is presented in Fig. 1, the nonlinear conductor being is plasma, *i.e.* a gaseous medium that contains free electrons, ions and atoms in the ground state. Its shape reveals that the gaseous conductor in a plasma diode evolves, when the voltage of the dc power supply connected to it is gradually increased and decreased, through a succession of abrupt variations of its current transport behavior. Such abrupt transitions can only arise in nonlinear systems that are driven away from equilibrium. These appear beyond a critical threshold when the system becomes unstable transiting into a new conduction state.

Applying a positive potential on an electrode playing the role of the anode in a device working as a plasma diode an electric field appears in front of it, which penetrates the gas on a certain distance. This distance depends on the voltage of the anode and on the plasma parameters. The electrons that "feel" the presence of the electric field are accelerated towards the anode. For values of the anode potential for which the kinetic energy of the electrons is not sufficient to produce excitations of the gas atoms the current collected by it increase linearly in agreement with Ohm's law (branch **0**-a). If the voltage of the dc power supply is gradually increased so that the electric field created by the anode accelerates the electrons at kinetic energies for which the atom excitation cross section suddenly increases, a part of electrons will lose their momentum. As a consequence, at a certain distance from the anode, which depends on its potential, a net negative space charge is formed. Acting as a barrier for the current its development simultaneously with the increase of the anode potential determines the appearance, in the current-voltage characteristic, of the branch **a-b** in which the current is nearly constant although the voltage on the gaseous conductor increases. The accumulation of electrons in a region where the translation symmetry of the excitation cross section, as a function of electron kinetic energy, is broken represents the first phase of spatial pattern formation in a plasma diode.

When the voltage of the dc power supply is additionally increased the electrons that have not lost their momentum after atom excitations will obtain the kinetic energy for which the ionization cross section suddenly increases. Because of the small differences of electrons' kinetic energy for which the excitation and ionization cross-sections suddenly increase, a relatively great number of positive ions are created adjacent to the region where electrons are accumulated. Since the electrons that ionized the neutrals, as well as those resulting from such a process, are quickly collected by the anode, a plasma enriched in positive ions appears between the net negative space charge and the anode. As a result of the electrostatic forces acting between the well-located negative space charge and the adjacent net positive space charge, the space charge configuration naturally evolves into an ordered spatial structure known as DL. This is located at a certain distance from the anode that depends in its potential. Its potential drop depends on the excitation and ionization rates sustained by the electrons accelerated towards the anode. The cause that produces the abrupt increase of the electrical conductance of the gaseous conductor emphasized in Fig. 1 by the sudden increase of the current when its density reaches a critical value (marked by **b** in Fig. 1) is the appearance of a first level of self-organization. This level of self-organization is revealed by the spontaneously self-assembly in front of the anode of a nearly spherical CSCC, the spatial stability of which being ensured by a DL. It appears when the correlation forces that act between the two adjacent opposite space charges located in front of the anode reach a critical value for which any small increase of the voltage of the dc power supply make the current to grow above the threshold value (marked by **b** in Fig.1) for which a first instability of the gaseous conductor starts. As already shown [2,3], this instability is related to a self-enhancement process of the production of positive ions that appears when the potential drop on the DL reaches a critical value. This critical value corresponds to the state of the DL for which the electrons accelerated within it produce by ionization, in a relatively small region, adjacent to the net negative space charge, a net positive charge able to balance it. The amount of positive ions created in this way in the vicinity of the net negative space charge increases the local potential drop in which the thermalized plasma electrons are accelerated. Because of electrons' kinetic energy dependence of the production rate of positive ions, a new amount of positive ions are added to the previously existent one. At its turn this determines a new increase of the local electric field in which the thermalized plasma electrons are accelerated and so on. After this positive feedback mechanism the region where the electrons are accumulated after excitation is shifted away from the anode. So the region where the concentration of positive ions grows suffers a displacement simultaneously with the region in which the net negative space charge is located. Owing to the fact that the production rate of positive ions is limited by the number of atoms present in the gaseous conductor when the pressure is maintained constant, the departure of the DL from the anode ceases at a certain distance from it. During this unstable phase the DL expands to the state for which the net negative space charge equilibrates the adjacent net positive space charge placed in its next vicinity but also that located in the nucleus of the CSCC. Taking into account that the expanding DL reveals a behavior similar to surfaces that separates two different media, *i.e.* surface tension behavior, the final shape of the DL corresponds to a state of local minimum of the free energy. Therefore the DL tends to evolve to a spherical fireball shape. However because the self-assembling mechanism of the DL requires a constant supply with electrons transported by the current, the CSCC must maintain the electrical contact with the anode. As a consequence, the CSCC appears in a plasma diode attached to the anode. Under such conditions the electrons that produce and result after ionizations in the nucleus of the CSCC are constantly collected by the anode, so that the flux of thermal electrons that pass through the DL from ensures the DL continuous self-assembly.

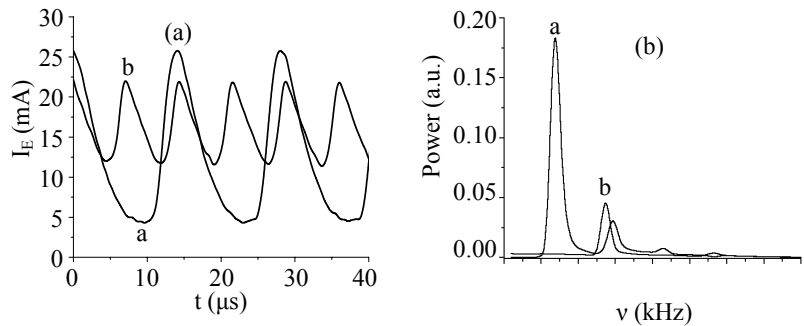

**Figure 2.** Oscillations generated by the S-shaped (corresponding to point **c** in Fig. 1) and Z-shaped bistabilities (corresponding to point **e** in Fig. 1); Power spectra for the oscillations shown in Fig. 2(a).

The abrupt increase of the current simultaneously observed with the emergence of the CSCC proves that the DL at its boundary accelerates electrons at kinetic energies for which positive ions are created. This signifies that the DL acts as an internal source of charged particles, the appearance of which increases the conductance of the gaseous conductor so that, for the same voltage of the anode, the current delivered by the dc power supply becomes greater. Since a nonlinear conductor is usually connected to the dc power supply through a load resistor the abrupt increase of the current is accompanied by a similar decrease of the potential supported by the conductor. This decrease of the potential drop on the conductor depends on the value of the load resistor.

The spatially ordered distribution of the opposite net electrical space charges in front of the anode displays all the characteristics of a spatial pattern formed by self-organization. Thus its emerge has the following characteristics. (i) It is self-assembled in a nonlinear system driven away from thermodynamical equilibrium when the external constraint (in our case the potential applied on the electrode) reaches a critical value. (ii) Its self-assembly process is initiated by spatial separation of net opposite space charges related to the symmetry breaking and spatial separation of the characteristic functions of the system (*i.e.* the neutral excitation and ionization cross sections as functions of the electron kinetic energy). (iii) The ordered structure self-assembles as a result of collective interactions between large groups of structural elements of the system (electrostatic attraction between opposite space charges). (iv) The groups of structural elements spontaneously emerge into a DL when the electrostatic forces, acting between them as long-range correlation, attain a critical value for which the space charge configuration naturally evolves into a state characterized by a local minimum of the free energy.

Although the self-assembly mechanism of the CSCC in front of the anode involves all key processes of a self-organized system this level of self-organization refers only to the emergence of a stationary spatial pattern. Another question related to this level of self-organization is the fact that the maintenance of the ordered spatial structure requires work performed by the dc power supply connected to the conductor. This work is done for ensuring the transport of thermalized plasma electrons towards the DL at the boundary of the stable CSCC. Only under such conditions the DL can ensure, by local acceleration of the thermalized plasma electrons, the physical phenomena required for its continuously self-assembling process. Therefore this state of self-organization is a metastable one because it requires work for maintaining stability.

In agreement with the above said the S-shaped bistability and its ability to work as a NDR has its origin in the described self-organization process during which matter and energy originated from the external dc power supply are stored in the CSCC. Since the self-assembling and de-aggregation processes of the CSCC sensitively depends on the current, fact demonstrated by the hysteresis loop **b-c-i-j-b**, it become possible to trigger these processes by oscillations that naturally appear when the gaseous conductor contains (or is connected to) a system able to perform natural oscillations. In such a case a sufficiently strong random variation of the current can stimulate oscillations when the voltage of the anode is placed in the voltage range where the gaseous conductor emphasizes an S-shaped bistability. Such current oscillations actually appear in a plasma device the current-voltage characteristic of which has the shape showed in Fig. 1. The shape of such oscillations obtained from a double plasma machine that works as a plasma diode are presented in Fig. 2. Their appearance was explained [4] revealing by experimental measurements that the CSCC has the ability to work as an S-shaped NDR and also as an oscillatory circuit. The reactive elements of this oscillatory circuit are the DL from its boundary, working as a capacitor, and an inductor, the physical origin of which is related to the differences in the masses of electrons and positive ions. The inductive behavior of the CSCC appears during its dynamics related to the periodic assembling and total (or partial) de-aggregation of the CSCC.

The S-shaped NDR displayed by the characteristic in Fig. 1 by the dashed line is particularly called current controlled NDR. As already showed it is associated to the self-assembling, in front of the anode, of a stable CSCC. Its genesis involves accumulation of charged particles and electric field energy, both of them provided by the external power supply. Since the self-assembling process of the CSCC sensitively depends on the current, it becomes possible to drive its formation and de-aggregation by changing the current. Therefore under conditions for which the CSCC reveals the behavior of a system able to perform natural oscillations the plasma device acts as a plasma oscillator. The appearance and maintenance of these oscillations is related to an aleatory process by which the potential of the anode is suddenly varied with a value for which the S-shaped bistability can work as an S-shaped NDR. These oscillations appear abruptly without revealing an amplitude growing process. Their appearance is an illustration of the way by which a saddle-node bifurcation appears in a plasma device. Their maintenance is ensured by an internal feedback mechanism by which the oscillations themselves drive the accumulation and the release of matter and energy related to the modification of the internal space charge configuration of the CSCC. Note that the CSCC actually reveals an internal "fine structure" [5] that makes it able to release only a part of the matter and energy accumulated during its self-assembling process. Therefore the S-shaped NDR can sustain oscillations by transiting from an unstable point placed on the line that represents the S-shaped NDR and the corresponding maximal value of the current. As known, the area of the hysteresis loop corresponds to the power extracted from the dc power supply during the CSCC self-assembling process.

As revealed by the I(V)-characteristic presented in Fig. 1, simultaneously with the emergence of the stable CSCC, the gaseous conductor transits into a state for which the same current is transported through the plasma for a smaller voltage applied on it. This reveals that the system "locked" in a state for which the power required to maintain it is ensured by a minimal "effort" from the external dc power supply. This minimal value of the power extracted from the dc power supply is related to the emergence of the stable CSCC acting as a new source of charged particles. It is located close to another minimum that appear when the voltage of the anode and, consequently, the current reaches the critical value marked by **d** in Fig. 1. For this value of the anode voltage the static current voltage characteristic shows the Z-shaped bistability. In that case the current decreases abruptly for the same value of the anode voltage. This new conduction state of the gaseous conductor for which the power required from the dc power supply becomes again minimal is related to the transition of the CSCC into the above-mentioned steady state. As shown in Fig. 2, simultaneously with this transition the oscillations generated by the plasma device have twice the frequency and half of the amplitude of the oscillations stimulated by the S-shaped NDR. Simultaneously with the sudden transition into this new conduction state the current collected by the anode becomes periodically limited. This phenomenon appreciated as the most celebrated diodic event [6] was explained [7] considering a new kind of instabilty that appears when the potential of the anode reaches the critical value for which the gaseous conductor reveals Z-shaped bistability. This bistability develops when the excitation and ionization rates at the two "ends" of the DL at the boundary of the stable CSCC become so high that the equilibrium between the two adjacent opposite space charges can be realized after ionization in a small region placed in the vicinity of the negative net space charge. Under such conditions the DL from the boundary of the CSCC becomes unstable because a small increase of the anode voltage that determines its departure from the anode initiates a new mechanism by which the DL is able to ensure its own existence. Thus, transiting into a moving state through a medium that contains thermalized electrons, namely the plasma, the DL becomes able to self-adjust its velocity at the value for which the additional flux of electrons transiting it is equal with the flux of electrons related to the decreasing of the current. Transiting into a moving phase the DL undertakes a work that diminishes the work required from the external dc power supply to maintain the new conduction state of the gaseous conductor.

As the genuine cause of the Z-shaped bistability it was identified the periodic limitation of the current related to the transition of the CSCC into a steady state during which the DL at its boundary is periodically detached and re-assembled [7]. Since after the departure of the DL from the boundary of the CSCC the conditions for the self-assembling of a new DL are present, an internal triggering mechanism appears, ensuring the periodicity of the consecutive formation, detachment and disruption of DLs from the CSCC boundary. Thus, after the departure of the DL from the anode the conditions for the self-assembling of a new DL in the same region where the first one was self-assembled, appear. The development of the negative space charge of the new DL acts as a barrier for the current, so that this is diminished at the critical value for which the existence conditions for the moving DL disappear. As a result, the moving DL disrupts. During the disruption process the electrons from the negative side of the DL becomes free moving as a bunch towards the anode. Reaching the region where the new DL is in the self-assembling phase the flux of electrons traversing it, and implicitly the ionization rate, suddenly increases at the value for which the new DL starts its moving phase. In this way an internal triggering mechanism ensures the successively self-assembly, detachment and de-aggregation of DLs from the boundary of the CSCC. The described dynamics of the DL related to the Z-shaped bistability of the gaseous conductor is experimentally proved by the appearance of oscillations with twice the frequency and half the amplitude of the oscillations sustained by the S-shaped bistability. This demonstrates that the dynamics of the successively self-assembled and de-aggregated DLs is produced in the time span corresponding to the period of the oscillations sustained by the S-shaped NDR.

We note that periodic limitation of the current related to the successive self-assembly and de-aggregation of DLs from the border of the CSCC requires a relatively small amount of matter and energy extracted from the dc power supply. Additionally, this periodical variation of the current does not represent genuine oscillations as those sustained in a resonant circuit by the S-shaped NDR. Therefore connecting an oscillatory circuit with a high quality factor to such kind of NDR, the oscillations reveal an amplitude growing process. The phenomenology related to the periodic limitation of the current is an illustrative model for the Hopf bifurcation.

## 3. Conclusions

The presented new information offered by plasma experiments concerning the actual physical basis of the S-shaped, respectively Z-shaped bistability and implicitly of the S-shaped and N-shaped NDRs reveal that two levels of self-organization are at their origin. The first level of self-organization is emphasized in the gaseous conductor by the emergence of a CSCC, the self-assembly and de-aggregation of which sensitively depend on the current collected by the anode. In that case the oscillations appear abruptly when to the S-shaped NDR is coupled a system able to perform natural oscillations.

The Z-shaped bistability and the current limiting phenomenon related to the N-shaped NDR are related to a higher level of self-organization emphasized by the presence of a CSCC able to ensure its existence in a steady state. During this steady state matter and energy are periodically exchanged between the CSCC and the environment. In this way the steady CSCC acts as the "vital" part of an oscillator, *i.e.* it stimulates and sustains oscillations in a system working as a resonator. In the last case the oscillations appear after an amplitude growing process.